\documentclass[12pt,preprint]{aastex}
\usepackage{emulateapj5}
\usepackage{epsfig}
\usepackage{graphicx}

\newcommand{\etal}{{\it et al.}}

\newcommand{\bq}{\begin{equation}}
\newcommand{\eq}{\end{equation}}
\newcommand{\msun}{\mbox{M$_{\odot}$}}

\newcommand{\eg}{{\it e.g.,}}
\newcommand{\ie}{{\it i.e.}}

\shorttitle{Early Universe Reionization}
\shortauthors{Panagia {\it et al.}}

\begin{document}

\title{Direct evidence for an early reionization of the
Universe?}\footnote{ Based on observations obtained with the NASA/ESA
Hubble Space Telescope obtained at the Space Telescope Science
Institute, which is operated by the Association of Universities for
Research in Astronomy (AURA), Inc., under NASA contract NAS5-26555; on
observations obtained at the European Southern Observatory (ESO) using
the ESO Very Large Telescope (VLT) on Cerro Paranal (Director's
Discretionary Time) and on observations made with the Spitzer Space
Telescope, which is operated by the Jet Propulsion Laboratory,
California Institute of Technology under NASA contract 1407.}

\author{N. Panagia\altaffilmark{2,3}, 
S.M. Fall\altaffilmark{2}, 
B. Mobasher\altaffilmark{2,3}, 
M. Dickinson\altaffilmark{4}, 
H.C. Ferguson\altaffilmark{2}, 
M. Giavalisco\altaffilmark{2},
D. Stern\altaffilmark{5}, 
T. Wiklind\altaffilmark{2,3}}
\altaffiltext{2}{Space Telescope Science Institute, 
3700 San Martin Drive, Baltimore, MD 21218, USA; panagia@stsci.edu,
fall@stsci.edu, mobasher@stsci.edu, ferguson@stsci.edu,
mauro@stsci.edu, wiklind@stsci.edu}
\altaffiltext{3}{Affiliated with the Space Sciences Department of the
European Space Agency}
\altaffiltext{4}{NOAO, 950 N. Cherry Ave., P.O. Box 26732, Tucson, AZ
85726-6732, USA; med@noao.edu}
\altaffiltext{5}{Jet Propulsion Laboratory, California Institute of
Technology, MS 169-506, Pasadena, CA 91109; stern@thisvi.jpl.nasa.gov}

\begin{abstract}
We examine the possible reionization of the intergalactic medium (IGM)
by the source UDF033238.7-274839.8 (hereafter HUDF-JD2), which was
discovered in deep {\it HST}/VLT/{\it Spitzer} images obtained as part
of the Great Observatory Origins Deep Survey and {\it Hubble}
Ultra-Deep Field  projects. Mobasher \etal~(2005) have identified
HUDF-JD2 as a massive ($\sim6\times10^{11}M_\odot$) post-starburst
galaxy at redshift z$\gtrsim6.5$.  We  find that HUDF-JD2  may be
capable of reionizing its surrounding region of the Universe, starting
the process at a redshift as high as z$\approx 15 \pm5$.

\end{abstract}

\keywords{galaxies: evolution ---  galaxies: high-redshift --- galaxies:
intergalactic medium --- galaxy: individual (HUDF-JD2) --- cosmology:
observations --- cosmology: theory --- cosmology: early universe}

\section{INTRODUCTION}

The reionization of the intergalactic medium (IGM) is undoubtedly one of
the most crucial events in the evolution of the Universe.  Since
reionization drastically reduced the opacity of the Universe to
ionizing radiation, it is likely  to  have influenced and/or determined
the formation and evolution of galaxies and other structures in the
early Universe (see, \eg~Loeb \& Barkana 2001). The presence of strong
Lyman-$\alpha$ forest absorption in the spectra of distant quasars indicates
that reionization was completed by $z \approx 6$ (Becker \etal~2001, Fan
\etal~2002), while the polarization of the cosmic
microwave background (CMB) radiation suggests it may have begun at a much
earlier epoch, $z \sim 17\pm 5$ (Spergel \etal~2003). Theoretical efforts to
understand and constrain the time evolution of cosmic reionization
sources  (\eg~Loeb \& Barkana 2001;  Stiavelli, Fall \& Panagia 2004a,
hereafter referred to as SFP) have to be confronted with observational
identifications of the ionization sources.  

The  Hubble Space Telescope (HST) observations obtained by the Great
Observatory Origins Deep Survey (GOODS) project (Giavalisco \etal~2004),
and more recently the observations of the  Hubble Ultra-Deep Field
(HUDF; Beckwith \etal~2005) have opened a window through which one may
hope to peek and see distant sources that could provide a direct clue to
the reionization process.  The available  data have already extended the
searches to $z\sim 6.5$ by  the drop-out techniques  (Dickinson
\etal~2004, Giavalisco \etal~2004, Bunker \etal~2004, Yan \& Windhorst
2004, Bouwens \etal~2004b, Eyles \etal~2005) and narrow-band surveys 
(Rhoads \etal~2004, Stern \etal~ 2005, Taniguchi \etal~ 2005). Yet
measures of the luminosity functions remain uncertain (Yan \& Windhorst
2004,  Bouwens \etal~2004a), and the identification of the reionization
sources is still just a possibility (\eg~Stiavelli, Fall \& Panagia
2004b).  This is because, even with the exquisite images of the HUDF, it
is hard to identify and characterize sources at redshifts appreciably
higher than 6, which are crucially important to properly define and
understand the processes associated with reionization.  

At present,  combined ultra-deep images of faint galaxies taken by the 
{\it Hubble Space Telescope} and {\it Spitzer Space Telescope} (Werner
\etal~2004) can be used to provide the multi-waveband information needed
to search for galaxies at even higher redshifts   and to explore their
nature. In particular,  the HUDF  observations
(Beckwith \etal~2005, Thompson \etal~2005), representing the deepest
images of the Universe at optical  ($B$, $V$, $i$, and $z$ bands;
HST/ACS) and near-infrared ($J$, $H$; HST/NICMOS) wavelengths,
combined with the VLT and Spitzer/IRAC observations obtained by the
GOODS project at NIR ($K_s$; ESO/VLT) and mid-infrared (3.6-24 $\mu$m; 
Spitzer/IRAC+MIPS) wavelengths (Vandame \etal~ 2005, in preparation,
Labb\'e \etal~2005, in preparation, Dickinson \etal~2005, in
preparation), are well suited to accomplish this aim. 
Analyzing these data, Mobasher \etal~(2005)  searched for  very red
galaxies with $(J-H)_{AB}>1.3$ and no detection at wavelengths shorter
than $J$ (\ie~$J-$band dropouts). They found two sources that  satisfy
the $J-$band dropout criteria, UDF033242.8-274809.4 and
UDF033238.7-274839.8 (hereafter HUDF-JD1 and HUDF-JD2, respectively). 
The two sources are well detected in the K-band (ISAAC). Both sources
are clearly detected in all four IRAC channels (3.6, 4.5, 5.8 and 8
$\mu$m). 

The first of theses sources, HUDF-JD1,  appears to be positionally
associated (within 1 arcsec) with an X-ray source.  Since no acceptable
fit to its observed spectral energy distribution (SED) could be found in
terms of stellar synthesis models, Mobasher \etal~(2005) conclude  that
it is most likely to be an obscured AGN  at intermediate redshift.

On the other hand, the presence of a clear Balmer break in the observed
SED of the HUDF-JD2 solidifies its high redshift identification and  
reveals a post-starburst population.   From their
best-fit models,  Mobasher \etal~(2005) derive a photometric redshift of
$z\sim6.5$ and a bolometric luminosity of L$_{\mathrm{bol}} = 1 \times
10^{12}$ L$_{\odot}$ (for a cosmology with $H_0=70$ km s$^{-1}$
Mpc$^{-1}$, $\Omega_{m}=0.3$ and $\Omega_{\Lambda}=0.7$).  Using the
M$_*$/L$_{\mathrm{bol}}$ ratio from the model fit, the mass in stars is
estimated to be  M$_* \simeq 6 \times 10^{11}$ M$_{\odot}$ for an
adopted Salpeter initial mass function  with minimum and maximum
stellar masses of 0.1 and 100~\msun, respectively.

The strong Balmer break displayed by HUDF-JD2 requires that its SED be
dominated by stars of spectral types A0 or later, \ie~main-sequence
stars  with masses less than $\sim3M_\odot$ (\eg~Allen 1973). This fact
unambiguously indicates a rather old age  ($>$300 Myr) for the stellar
population of HUDF-JD2. Mobasher \etal~(2005) conclude that the stars
were formed at z$>9$ (\ie~an age of the Universe $t_{U}< 540$ Myr), and
possibly as high as z$\sim$12-20, so that most of the stars were in
place when the Universe was only $\sim 200-400$ Myr old.  Moreover,
their model fitting indicates that HUDF-JD2 formed the bulk of its stars
very rapidly, on time scales $\leq 100$ Myr, and the subsequent evolution
was essentially passive. 
A stringent upper limit to the starburst age is set by the photometric
redshift $z\simeq 6.5$ of HUDF-JD2, which corresponds to a time
when the Universe was only 830 Myr old.  Adopting this prior, the  model
fitting procedure indicates that the age of the stellar population in
HUDF-JD2 is likely to be bracketed between 350 and 650 Myr.  These
ages correspond to redshifts of galaxy formation between 10 and 20.

The estimated reddening is quite modest, with $E(B-V)\leq 0.06$ at 95\%\
confidence limit. The best-fit model SED also provides estimates of the
overall metallicity of the galaxy, which is  bracketed within the
interval $Z=0.02-0.004$, \ie~between $Z_\odot$ and $Z_\odot/5$. In the
following, we will use a fiducial value of  $Z=0.008=0.4Z_\odot$,
keeping in mind that this value may be  uncertain by a factor of two. 

Given the high redshift of HUDF-JD2 and the short time scale over which
its stellar population was formed, it is of interest to explore whether
the radiation from this galaxy was able to reionize its surrounding
intergalactic medium. In this letter, we estimate the output of ionizing
radiation from HUDF-JD2 over its past history. We will show that the
Lyman continuum flux radiated by this source in the early phases of its
evolution may be enough to reionize its surrounding IGM  appreciably
and, under favorable conditions, even ionize it completely.  For
consistency with Mobasher \etal~(2005) paper,  we adopt  $H_0=70$ km
s$^{-1}$ Mpc$^{-1}$, $\Omega_{m}=0.3$ and $\Omega_{\Lambda}=0.7$.

\section{IMPLICATIONS FOR REIONIZATION}

Reionization is a process that depends on the UV output of  galaxies
integrated over the  time interval when they are active UV sources.   In
this respect, HUDF-JD2 is a promising reionization source in that, 
according to the models, its integrated Lyman continuum photon flux is
about $4\times10^{72}$ photons  that had to ionize the neutral hydrogen
gas within a comoving volume of about $14-18\times10^4$ Mpc$^3$, the
exact volume depending on the specific redshift (within the range
$z\sim9-15$; see below) at which H was fully ionized.   For a comoving hydrogen
density of $6.2\times10^{66}$ atoms~Mpc$^{-3}$ (\eg~Spergel \etal~2003),
it appears that the produced Lyman continuum photons outnumber the  H
atoms by almost an order of magnitude and that, therefore, a complete
reionization is a real possibility.  
However, a realistic treatment of this process should take into account
the facts that only a fraction $f$ of the Lyman continuum photons 
escape from the source, and that the ionized gas may recombine, and do
so at a particularly high rate if  the IGM material is clumped,
traditionally parametrized by the clumping factor,  
$C=<n{^2_H}>/<n_H>^2$. These effects tend to reduce the overall
efficiency of the ionization process. 

Here,  we take advantage of the extensive set of model calculations made
by SFP who considered a wide range of  semi-analytical 
models to estimate the reionization of the
IGM.  All cases studied by SFP were based on the implicit
assumption of a ``snapshot" of a statistical assembly of cosmic sources,
which possibly formed at different epochs and were  efficient UV
radiation emitters over different time intervals, but, {\it on average},
constituted a steady supply of ionizing photons over the chosen redshift
interval.  The situation studied here is quite different. HUDF-JD2 is 
a single, powerful source for which both the birth time and 
subsequent evolution are constrained by the observed SED. Therefore,  
we have to allow explicitly  for its past
evolution.  

Although the physical quantity that determines the reionization is the
total (\ie~time integrated) Lyman continuum photon flux, for a
convenient comparison with observations, SFP  parametrize their models
in terms of the ratio  of the  Lyman-continuum photon flux to the UV
flux at 1400\AA, $\epsilon_{ion} = N_{Lyc}/(\nu L_\nu)_{1400}$, which 
by construction  remains constant over the time reionization takes
place. For the case of HUDF-JD2, it is convenient  to consider the
ratio of a {\it time averaged}  Lyman-continuum photon flux to
the {\it current} UV flux at 1400\AA.  First, we define the average
Lyman-continuum photon flux as:

\begin{equation}
<N_{Lyc}(t)> = \frac {\int_{t_1}^{t} N_{Lyc}(t) dt} {min[(t-t_1), \Delta
t_{reion}]}
\end{equation}

\noindent where $N_{Lyc}(t)$ is the instantaneous Lyman-continuum photon flux
of the galaxy, $t_1$ is the time of the galaxy birth, and $\Delta
t_{reion}$ is the time interval  during which reionization takes place.
We adopt $\Delta t_{reion}= 100$~Myr, which coincides with the
approximate  duration of the starburst in HUDF-JD2 as estimated by
Mobasher \etal (2005).  We can relate the average  Lyman-continuum
photon flux $<N_{Lyc}(t)>$ to the luminosity in the detectable UV  by
defining an  effective ionization parameter:

\begin{equation}
\epsilon{^{eff}_{ion}}(t) = \frac {<N_{Lyc}(t)>}{(\nu L_\nu(t))_{1400}}
\end{equation}

Figure 1 shows the  ionization parameter as a function of age for a
starburst lasting 100 Myr, which is computed using Starburst99 models
(Leitherer \etal~1999) for different  metallicities and adopting a
Salpeter initial mass function within the interval 1 and 100 M$_\odot$. 
For our range of metallicity, the value of
$\epsilon_{ion}$ at about 350-650~Myrs (\ie~the range within which the
HUDF-JD2 age is likely to lie) is rather high. For our fiducial
metallicity of $0.4Z_\odot=0.008$ and for an age of 500~Myr, we find
$\epsilon{^{eff}_{ion}}=9.6\times 10^{11}$~erg$^{-1}$, \ie~4.8 times
higher than the value for a 100,000K blackbody source,  even though in
the synthesis models no star ever attains such a high effective
temperature.  This is because the ionizing flux was very high at the
early epochs, while the observed 1400\AA~ flux correponds to a later
epoch when the starburst has weakened considerably.  We note that since
the ionization parameter  depends rather strongly on the age and
metallicity of the galaxy, $\epsilon{^{eff}_{ion}}$  may be uncertain by
a factor of two. The results are nearly the same for any starburst
duration shorter than 100~Myr.   In this context, we note that a
relatively long starburst (\ie~duration $\sim$100~Myr) is probably
needed to account for the high metallicities implied by the best-fitting
SED and that these, in turn, may require multiple episodes  of nuclear
reprocessing.

Scaling the results of SFP  model calculations for $10^5\ K$ (their
Table~1, and Appendix A), we have computed the constraints on source
surface brightness for a  case in which reionization starts at $z=13$
and is complete by $z=10.7$ (\ie~about 500 and 400 Myr, respectively,  before the epoch
corresponding to our observations) with ionization efficiency
$\epsilon{^{eff}_{ion}}=9.6\times 10^{11}$~erg$^{-1}$ (see Figure 2). In
this diagram the scaling factor is simply the ratio of the 
$\epsilon{^{eff}_{ion}}$ value to the reference value of SFP for
$T_{BB}=100,000$K, $\epsilon_{ion}(100,000K)=2\times
10^{11}$~erg$^{-1}$. The one difference relative to SFP model calculations
is that here the only contribution to the near UV flux is from stellar
radiation because the nebular  emission of the galaxy is
negligible at the time it is observed. This simple scaling procedure can
be  employed to describe all cases involving individual sources  with
well defined evolutionary histories. 

In our case,  we find that with the  observed magnitude at
$\lambda\simeq 1.1 \mu m$ (corresponding to rest-frame wavelength
1400\AA), $m_{AB}\simeq 27.0$, and an effective source density of $N\sim
1/(2.5)^2= 0.16$ arcmin$^{-2}$ (\ie~one source within  the NICMOS field,
$2.5'\times 2.5'$), HUDF-JD2, as an isolated source, could have ionized
the hydrogen gas within the volume encompassed within the solid angle
subtended by the HUDF-NICMOS boundaries up to the redshift $z\sim10.7$
(\ie~the  {\it ``cell of Universe"} that we can associate to it) only if
the escape fraction of Lyman-continuum photons is $f>0.26$ (for C=1) or
the clumping factor is  $C< 21$ (for $f=1$) (see Figure 2). This
is already an impressive result because  we are dealing with  {\it one}
source, which, even with  rather severe assumptions, such as $f\sim0.2$
and $C\sim20$, can be responsible for at least 20\% of the ionization of
its Universe cell.  

By taking an escape fraction less than unity we are implicitly including
the effects of absorption of Lyman continuum photons by both neutral
hydrogen and dust  grains within HUDF-JD2. However, in our calculations
we also have to allow for dust extinction that may attenuate the source
spectrum at wavelengths longer than the Lyman limit.  The best-fit to 
HUDF-JD2 SED yields an  estimate of  a rather modest selective
extinction $E(B-V)\lesssim 0.06$ (Mobasher \etal~ 2005), which implies
an upper limit to the extinction at rest frame 1400\AA~  of about $0.54$
magnitudes. This result is retatively independent of the adopted 
extinction law, be it the effective attenuation law for starburts
galaxies (\eg~Calzetti  \etal~ 1996,  Charlot \& Fall 2000),  or the
recent measurements of the  extinction law in a galaxy at $z=0.83 $
(Mediavilla \etal~2005).  The extinction correction increases the source
flux at rest frame 1400\AA~  by a factor of
$\sim10^{0.4\times0.54/2}=1.28$ on average, thus moving the source to
the right in Figure 2 (open triangle), indicating  that for $f\sim0.2$
and $C\sim20$HUDF-JD2 by itself  could  account for up to 25\% of the
total ionization of its surrounding IGM.

In addition, HUDF-JD2 may not be an isolated source, but rather the
brightest member of a population of galaxies within one and the same
volume, whose other members are too faint to be detected in the
available images. Adopting a Schechter luminosity function with  a slope
$\alpha=1.6$ and a cutoff magnitude $M^*_{UV}=-21.2$ (as appropriate for
z=3 galaxies; Steidel \etal~1999), with the condition that  HUDF-JD2 is
the only source brighter than $m_{AB}\simeq27$ or $M_{AB}\sim-21.0$, and
assuming that the other members of the same group are identical in
nature to HUDF-JD2 except for being fainter, the total power turns out
to be between 4.2 and 6.7 times higher than HUDF-JD2 itself, depending
on whether the LF extends just to $L^*/10$ or down to $L^*/1000$,
respectively.  As a consequence, the global effect for reionization
would correspond to that of a source density $\sim 5.4\pm1.2$ higher,
\ie~0.86~arcmin$^{-2}$ (open circle in Figure 2).  If this is the case,
then it follows that HUDF-JD2, together with its lower luminosity
companions, could {\it easily} produce enough Lyman-continuum radiation
to reionize the IGM in  their neighborhood under most reasonable
circumstances ({\it e.g.}, escape fraction as low as 0.2 and clumping
factor as high as 20).  Correspondingly the  electron scattering optical
depth for our reference model, in which reionization starts at z$_1=13$
and is complete at z$_2=10.7$, turns out to be $\tau_e=0.13$.

\section{DISCUSSION AND CONCLUSIONS}

We have considered the possible effects  the source HUDF-JD2 on the
reionization of the Universe. We find that HUDF-JD2  may indeed  be
capable of reionizing its portion of IGM (\ie~the gas within a volume 
defined by the subtended solid angle of the HUDF-NICMOS field and the
redshift of complete reionization, z$_2\sim9-15$),  possibly with the
help of unseen fainter companions, starting the process at a  redshift
as high as z$_1\simeq 15 \pm 5$.

Regardless of whether HUDF-JD2 is able to account for the
{\it entire} reionization of its cell of Universe, it is certain that it
had the power to appreciably ionize it at high redshifts. Even with no
reddening correction and not allowing for faint companions, it appears
that  HUDF-JD2 alone should be able to ionize either 20\% of its cell if
the matter distribution is homogeneous, or ionize it at an average  20\%
level (\ie~the gas in the cell is 80\% neutral) if there are density
inhomogeneities in different directions .  Mobasher \etal~(2005) estimate
that active star formation in HUDF-JD2 lasted less than 100~Myr, which
implies a redshift interval of $\Delta z\sim 2$ if the galaxy formed at
$z\sim12$ and  $\Delta z\sim6$ if the galaxy formed at $z\sim20$, 
during which HUDF-JD2 was ionizing its cell of the Universe at a $\sim
20$\% level, at least.  More generally, if HUDF-JD2 represents the
``typical" source present in {\it any} $2.5'\times2.5'$ field on the
sky, then  this type of sources could readily ionize the IGM  at 
levels and redshifts as high as required to account for WMAP findings
($\tau_e=0.17\pm0.06$ at z$=17\pm5$; Spergel \etal~ 2003).

It is interesting to note that the four $i'$-dropout sources detected
and studied by Eyles \etal~ (2005) in the HST/ACS GOODS images of the
Chandra Deep Field South, may indeed be representatives of  the ``second
brightest" galaxies in the ladder of the luminosity function. Actually,
their measured magnitudes are about 1.5-2 mag fainter than measured for
HUD-JD2.  All of these sources are found to have z$\sim5.8$, stellar
masses  $2-4\times 10^{10}\msun$, ages of $250-650$ Myr and implied
formation redshifts $z_f\sim7.5-13.5$ (Eyles \etal~ 2005).  The presence
of HUD-JD2 together with these additional bright sources in the Chandra 
Deep Field South (by design one of them, namely SMB03\#1, falls within
the HUDF area) appears to lend support to the possibility that the
reionization of the Universe be dominated by massive galaxies that,
rather unexpectedly, were able to form  during the early evolution of
the Universe.

Although the presence of a cluster of galaxies and the assumptions about
their properties are somewhat speculative, we would like to stress that
these are {\it predictions} that will be possible to confirm or exclude
once the James Webb Space Telescope (JWST; \eg~Stiavelli \etal~ 2004),
currently due to be launched in mid-2013, will come into  operation.  
According to the model calculations by Panagia \etal~(2003, and in
preparation), sources like HUDF-JD2 and its analogs at higher redshifts
will be easy to study spectroscopically because in their  active star
formation phases, they will have intrinsically high luminosities
providing fluxes as high as 150~nJy in the near and mid infrared. 
Fluxes at these levels are expected to be detectable with the
JWST-NIRSpec spectograph at resolution of $\sim1000$ with a S/N ratio of
10 in 10$^5$ seconds exposures.  With the advent of JWST it will be
possible to fully characterize these early Universe sources, thus
clarifying once and for all the process of galaxy formation and the
reionization of the Universe.

\section{ACKNOWLEDGMENTS}

We wish to thank Massimo Stiavelli, Anton Koekemoer,  and Mario Livio
for valuable discussions. The comments of an anonymous referee were
useful to improve the presentation of this paper. Support for the
GOODS-ACS program was provided  by NASA through a grant from the Space
Telescope Science Institute, which is operated by the Association of
Universities for Research in Astronomy, Inc., under NASA contract NAS
5-26555. Support for this work was also provided by NASA  through
Contract Number 1224666 issued by issued by the Jet Propulsion
Laboratory, California Institute of Technology, under NASA contract
1407. The work of DS was carried out at the Jet Propulsion Laboratory,
California Institute of Technology, under contract with NASA.

\clearpage

\begin{figure}
\epsscale{0.8}
\plotone{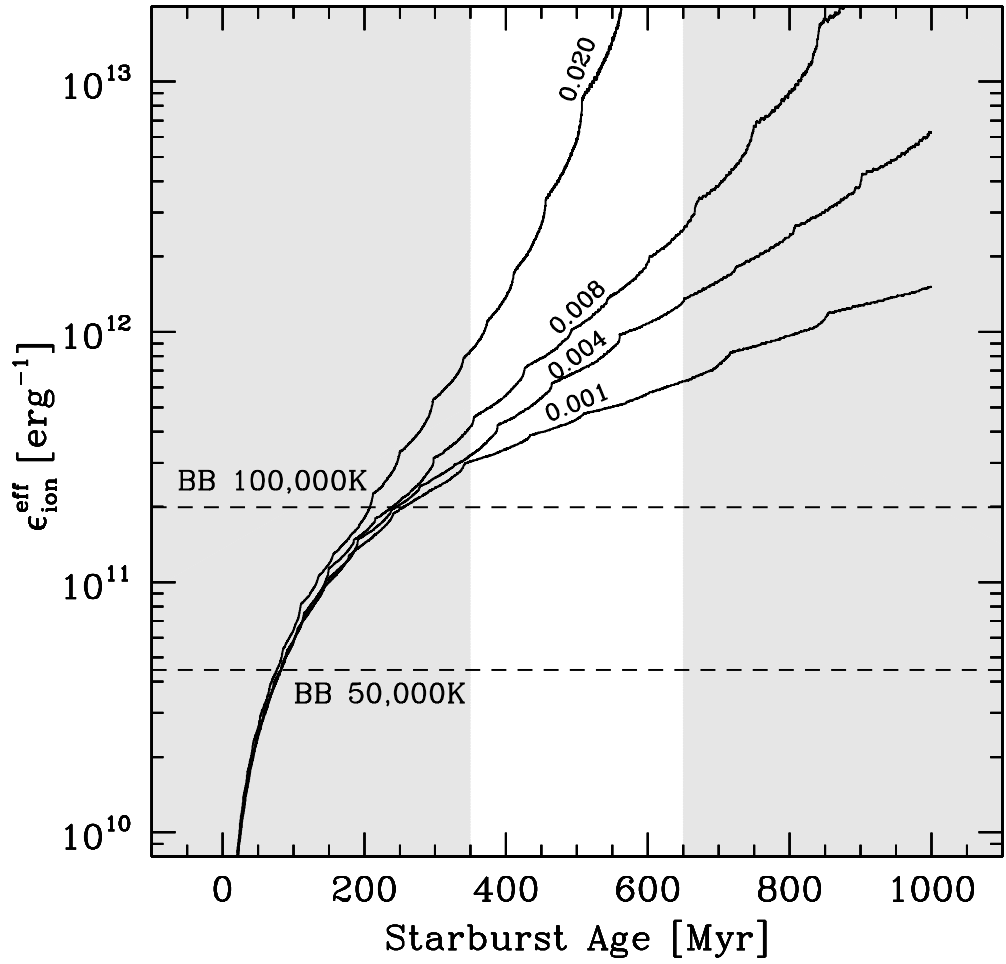}
\caption{Effective ionization parameter
$\epsilon{^{eff}_{ion}} = <N_{Lyc}>/(\nu L_\nu)_{1400}$ as a function of
starburst age  for different metallicities (indicated as metal
abundances by mass).  For comparison, the values of the ionization
parameters for black-bodies with temperatures of 50,000 and 100,000~K are also
shown. The unshaded area corresponds to the interval within which
the age of HUDF-JD2 is likely to lie.}
\label{fig1}
\end{figure}

\begin{figure}
\epsscale{0.8}
\plotone{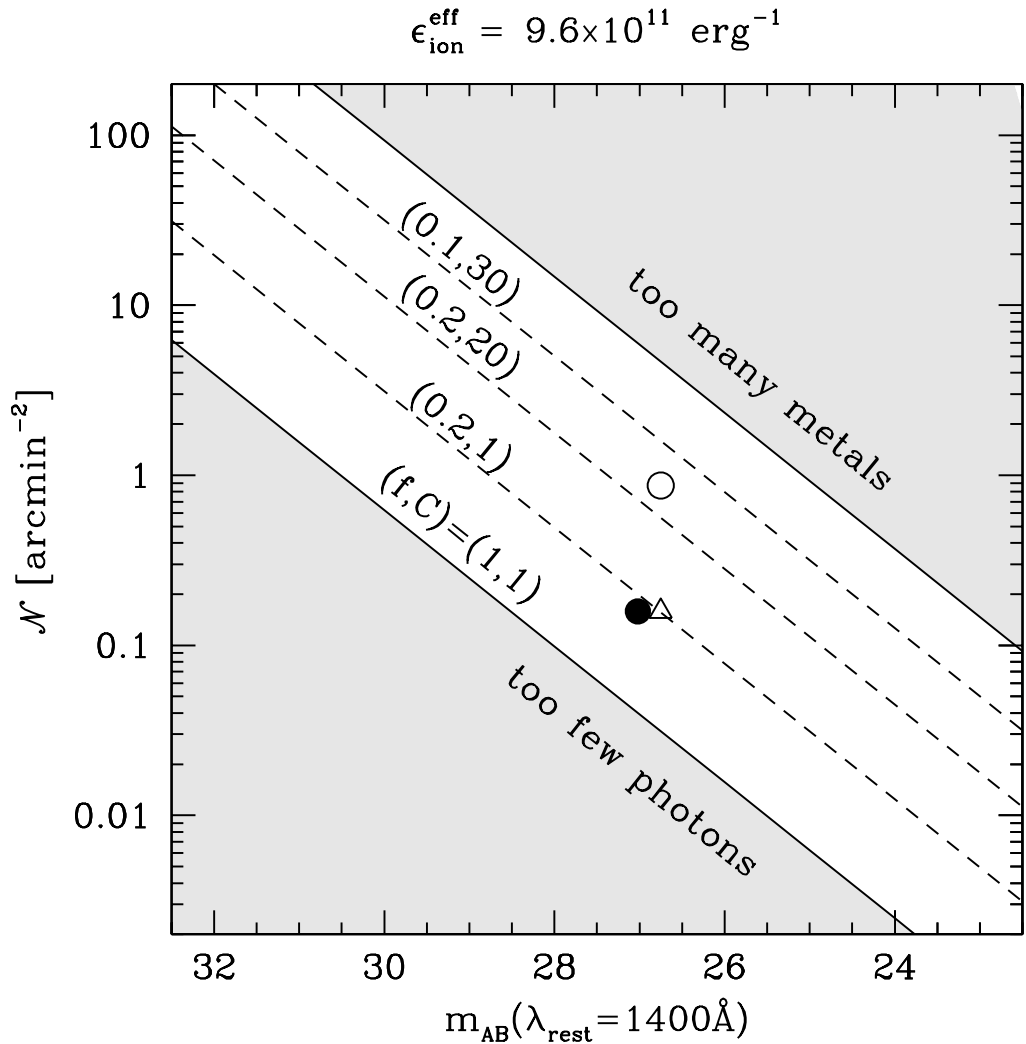}
\caption{Surface density of possible reionization sources as a function
of the apparent AB magnitude for identical sources with ionization
parameter $\epsilon{^{eff}_{ion}} = 9.6\times 10^{11}$~erg$^{-1}$. The
lower solid line represents the minimum surface brightness model, in
which the ionizing radiation escape fraction $f$ and the clumping factor
$C$ are both unity,  while the upper solid line represents the global
metallicity constraint $Z < 0.01 Z_\odot$ at z = 6.5 (see SFP). The
dashed lines represent models with $(f,C)=(0.2,1)$, $(0.2,20)$, and
$(0.1,30)$, respectively, characterized by a lower  ionizing
efficiency. The position of  HUDF-JD2 on the basis of its observed
fluxes is denoted with a dot, whereas the triangle and the circle denote
the positions after correction for dust extinction and after allowance
for unseen companions, respectively (see text).}
\label{fig2}
\end{figure}

\end{document}